\documentclass[showpacs,prb]{revtex4}
\usepackage{bm}
\usepackage{graphicx}
\usepackage{amsmath}
\usepackage{eufrak}
\usepackage{color}
\newcommand{\nix}[1]{}
\begin{document}

\title{Spin and valley-orbit splittings in SiGe/Si
heterostructures}
\author{M.O.~Nestoklon}
\author{L.E.~Golub}
\author{E.L.~Ivchenko}

\affiliation{A.F.~Ioffe Physico-Technical Institute, Russian
Academy of Sciences, St. Petersburg 194021, Russia}
\begin{abstract}
Spin and valley-orbit splittings are calculated in SiGe/Si/SiGe quantum wells (QWs) by using the tight-binding approach. In accordance with the symmetry considerations an existence of spin splitting of electronic states in perfect QWs with an odd number of Si atomic planes is microscopically demonstrated. The spin splitting oscillates with QW width and these oscillations related to the inter-valley reflection of an electron wave from the interfaces. It is shown that the splittings under study can efficiently  be described by an extended envelope-function approach 
taking into account the spin- and valley-dependent interface mixing. 
The obtained results provide a theoretical base to the experimentally observed 
electron spin relaxation times in SiGe/Si/SiGe QWs.
\end{abstract}
\maketitle

\section{Introduction}
At present various semiconductor materials are being involved in
the spintronics activities. SiGe/Si quantum well (QW) structures
are among them. Silicon-based systems can be particularly promising
due to a comparatively weak spin-orbit interaction and long electron
spin-relaxation times. Although bulk Si and Ge have an inversion
center, QW structures grown from these materials can lack such a
center and allow the spin splitting of the electronic subbands,
even in the absence of structure inversion
asymmetry.\cite{Golub_Ivchenko} An ideal SiGe/Si/SiGe QW
structure with an odd number of Si atomic planes is characterized
by the D$_{2d}$ point-group symmetry and, therefore, allows
spin-dependent linear-in-${\bm k}$ terms in the electron effective
Hamiltonian
\begin{equation} \label{1}
{\cal H}^{(1)}({\bm k}_{\parallel}) = \alpha (\sigma_x k_x - \sigma_y k_y)\:,
\end{equation}
where $\sigma_x, \sigma_y$ are the spin Pauli matrices, ${\bm k}_{\parallel}$
is the two-dimensional wave vector with the in-plane components
$k_x, k_y$, and $x \parallel [100], y \parallel [010]$.

\begin{figure}[b]
  \centering
    \includegraphics[width=.7\textwidth]{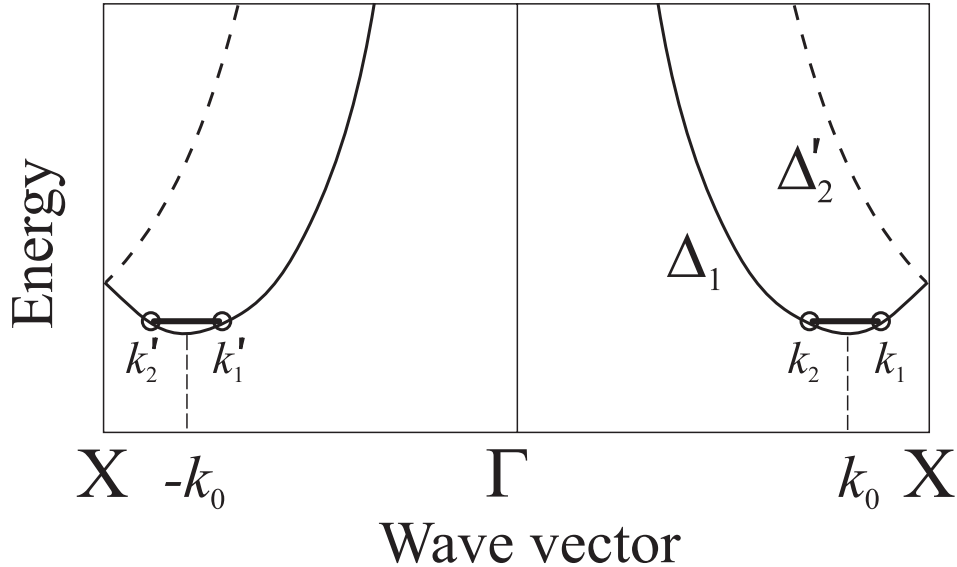}
  \caption{Schematic representation of the lower conduction
bands $\Delta_1$ (solid curve) and $\Delta'_2$ (dashed) in bulk Si
along the $\Gamma$--$X$ direction in the first Brillouin zone.
Horizontal bars in figure illustrate extension of the $e1$
quantum-confined state in the ${\bm k}$ space; ${\bm k}_1$,
${\bm k}_2$, ${\bm k}_1'$ and ${\bm k}_2'$ are wave vectors of
four Bloch states mixed in a QW.}\label{f_valleys}
\end{figure}

In the present work we use both the microscopic tight-binding
model and the envelope-function approach to calculate the spin
splitting of the conduction subbands in diamond-lattice QWs. The
obtained results are of particular interest in connection with the
experimental studies of electron spin relaxation in Si/SiGe
heterostructures.\cite{Wilamowski,Tyryshkin} The
consideration of a Si/SiGe structure with perfect interfaces and
without built-in electric fields allows one to put the upper limit
to the electron spin relaxation time.

In a bulk homogeneous sample of Si, two of six equivalent minima
of the conduction band $\Delta_1$ are located in two points, ${\bm
k}_0$ and $- {\bm k}_0$, along the direction [001] of the first
Brillouin zone as illustrated in Fig.~\ref{f_valleys}. The point-group
symmetry of a Si/SiGe(001) QW reduces~\cite{Golub_Ivchenko} and allows mixing between four bulk Bloch states attached to the ${\bm k}_0$ and $- {\bm k}_0$
valleys.\cite{Ohkawa,Boykin_splitting,Boykin_split2} The
valley-orbit mixing occurs under electron reflection from a
heterointerface: an electron with the wave vector ${\bm k}_1
\approx {\bm k}_0$ is reflected not only to the state ${\bm k}_2 $
attached to the same valley ${\bm k}_0$ but also to the state
${\bm k}_2'$ in the second valley $- {\bm k}_0$, see Fig.~\ref{f_valleys}.
The reflected wave is a superposition of two waves with their phase difference
dependent on the distance $z$ from the interface as $2k_0z$.
In the QW grown along the [001] direction,
quantum-confined electron states are standing waves formed as a result
of multiple reflection of the four waves ${\bm k}_1$, ${\bm k}_2$, ${\bm k}_1'$, ${\bm k}_2'$, or $\pm {\bm k}_0 \pm ({\bm k}_1 - {\bm k}_0)$, from the both heterointerfaces.

The spin splitting in conduction subbands is directly related to
spin dependence of the electron oblique-incidence reflection from
an interface. Spin-dependent reflection of an electron wave from
interface consists of intra- and intervalley contributions. The
latter should oscillate with the QW width $L$ in the same way as
the spin-independent valley-orbit splitting. Thus,
interface-induced spin splitting $\Delta_{\rm{spin}}$ contains two
contributions one oscillating with $L$ and another being smooth.
Their relation can be obtained in microscopic evaluations.

The paper is organized as follows. In Sec.~II we extend the
envelope function method to take into account intra- and
inter-valley spin-dependent contributions to the effective
interface potential. In Sec.~III we develop the $sp^3s^*$
tight-binding model in order to calculate the dependence of the
coefficient $\alpha$ in Eq.~(\ref{1}) on the QW width, discuss the
results of calculations and compare them with the analytical
equations derived in Sec.~II. The paper is concluded by Sec.~IV.
\section{Extended envelope function method}
Let us consider a QW layer A sandwiched between
barriers B and C on the right- and left-hand sides, respectively.
We assume that the three bulk materials $j =$ A, B, C have the
diamond-like lattice, the structure is grown along the principal
crystallographic axis $z \parallel [001]$, and the lowest
conduction subband $e1$ is formed by electronic states in the two
$\Delta$ valleys with the extremum points $\pm {\bm k}_{0j} =
(0,0, \pm k_{0j})$. Note that, in the Si$_{1-x}$Ge$_x$ solid
solution, the extremum-point position is a function of the content
$x$ and values of $k_{0j}$ are layer dependent. Because of the
lattice constant mismatch some of the structure layers are
strained. The layers B and C are assumed to be thick enough for
the tunnelling tails of the quantum-confined $e1$ states to decay
within these layers so that they can be considered as semi-infinite.

In the generalized envelope function approximation the electron
wave function $\Psi({\bf r})$ inside the layer $j$ is written as
\begin{equation} \label{psi}
\Psi({\bm r}) =  {\rm e}^{{\rm i} {\bm k}_{\parallel} \cdot {\bm \rho}}
[ \varphi_1(z; j) \psi_{ {\bm k}_{0j} }({\bm r}) + \varphi_2(z; j)
\psi_{- {\bm k}_{0j} }({\bm r})]\:.
\end{equation}
Here
\begin{equation} \label{Bloch}
\psi_{{\bm k}_{0j}}({\bm r}) = {\rm e}^{ {\rm i} k_{0j} z} u_{{\bm
k}_{0j}}({\bm r})\hspace{7 mm} \mbox{and} \hspace{7 mm}
\psi_{-{\bm k}_{0j}}({\bm r}) = {\rm e}^{ - {\rm i} k_{0j} z} u_{-
{\bm k}_{0j}}({\bm r})
\end{equation}
are the scalar Bloch functions at the two $\Delta$ extremum
points, $u_{\pm{\bm k}_{0j}}({\bm r})$ are the Bloch periodic
amplitudes, $\varphi_1(z; j)$ and $\varphi_2(z; j)$ are the smooth
spinor envelope functions defined within the layer $j$,
$\bm{\rho}$ is the in-plane component of the three-dimensional
radius-vector ${\bm r}$.

The two-valley effective Hamiltonian ${\cal H}$ is presented as a
sum of the zero-order valley- and spin-independent term
\begin{equation} \label{h0}
{\cal H}_0 = \frac{\hbar^2}{2} \left[ -  \frac{d}{dz}
\frac{1}{m_l(z)} \frac{d}{dz} + \frac{k_x^2 + k_y^2}{m_t(z)}
\right]
\end{equation}
and the interface-induced $\delta$-functional perturbation
\begin{equation} \label{h1}
{\cal H}' = V_L \delta(z - z_L) + V_R \delta(z - z_R)\:.
\end{equation}
Here $m_l$ and $m_t$ are the longitudinal and transverse effective
masses for electrons in the $\Delta$ valley, $z_L$ and $z_R$ are
the coordinates of the left- and right-hand side interfaces, $V_L$
and $V_R$ are both valley- and spin-dependent operators. Hereafter
we assume that the latter contain no differentiation $d/dz$, this
assumption excludes the need in symmetrization of $V_{L, R}$ and
the $\delta$-function.

The form of $V_L, V_R$ can be specified by applying the symmetry
considerations. A single (001) interface is characterized by the
C$_{2v}$ point-group symmetry allowing two linear-${\bm
k}_{\parallel}$ spin-dependent invariants, namely,
$$
h({\bm k}) = \sigma_x k_x -\sigma_y k_y \hspace{8 mm} \mbox{and}
\hspace{8 mm} h'({\bm k}) = \sigma_x k_y - \sigma_y k_x\:.
$$
It follows then
that the matrices $V_m$ ($m = L, R$) acting on the bispinor vector
$(\varphi_{1, 1/2}, \varphi_{1, -1/2},\varphi_{2, 1/2},\varphi_{2,
-1/2})$ can be presented in the form of a 2$\times$2 block matrix
\begin{equation} \label{2valley}
V_m = \left[ \begin{array}{cc} S_m h({\bm k}) + S'_m h'({\bm k}) &
\Lambda_m I + P_m h({\bm k}) + P'_m h'({\bm k}) \\
\Lambda_m^* I + P^*_m h({\bm k}) + P_m^{\prime *} h'({\bm
k}) & S_m h({\bm k}) + S'_m h'({\bm k})
\end{array} \right]
\end{equation}
with its components being linear combinations of the Pauli
matrices and the 2$\times$2 unit matrix $I$. Here ${\bm k} \equiv
{\bm k}_{\parallel}$, and $S_m, S'_m, \Lambda_m, P_m, P'_m$ are
coefficients characterizing the right-hand ($m=R$) and left-hand
($m=L$) interfaces, the first two of them ($S_m$, $S'_m$) are real
while others are complex. The diagonal components $V_{m; 11} =
V_{m; 22}$ give intra-valley contributions whereas the
off-diagonal components $V_{m; 12} = V^{\dag}_{m; 21}$ describe
interface-induced inter-valley mixing. It is more convenient to
perform the further considerations for a particular case of
coinciding barriers, C = B, and coinciding extremum points, ${\bm
k}_{0 {\rm B} } = {\bm k}_{0 {\rm A} }$ (or $k_{0{\rm B}} =
k_{0{\rm A}}$). Then we briefly discuss how these considerations
are generalized with allowance for C $\neq$ B and different
positions of extremum points ${\bm k}_{0j}$.

The choice of the electron Hamiltonian in the form of
Eqs.~(\ref{h0}), (\ref{h1}) corresponds to a particular set of
boundary conditions. For the structure B/A/B with $k_{0{\rm B}} =
k_{0{\rm A}} \equiv k_0$, this set reads
\[
\varphi(z_L+0) = \varphi(z_L-0)\:,
\:\varphi(z_R+0) = \varphi(z_R-0)\:,
\]

\begin{equation} \label{BC}
\frac{1}{m_l({\rm B})} \left( \frac{d \varphi}{dz}
\right)_{z_L-0}= \frac{1}{m_l({\rm A})} \left( \frac{d
\varphi}{dz} \right)_{z_L+0} + \frac{2}{\hbar^2}\ V_L\
\varphi(z_L)\:,
\end{equation}
\[
\frac{1}{m_l({\rm B})} \left( \frac{d \varphi}{dz}
\right)_{z_R+0}= \frac{1}{m_l({\rm A})} \left( \frac{d
\varphi}{dz} \right)_{z_R-0} - \frac{2}{\hbar^2}\ V_R\
\varphi(z_R)\:,
\]
where $\varphi(z_{L,R}\pm0)$, $(d \varphi/dz)_{z_{L,R}\pm0}$ are
the envelope function and its first derivative at $z$ approaching
the interface $L,R$ from the right- ($+0$) and left-hand ($-0$)
sides.

The next step is to analyze the phases of the coefficients $\Lambda_m, 
P_m, P'_m$ in the off-diagonal components of $V_m$ and establish a 
relation between $V_{\rm L}$ and $V_{\rm R}$. First of all, we take into
account that the translation of the radius-vector ${\bm r}$ by a
three-dimensional Bravais-lattice vector ${\bm a}$, results in a
multiplication of the Bloch functions $\psi_{\pm {\bm
k}_{0j}}({\bm r})$ in Eq.~(\ref{Bloch}) by the factors $
\exp{(\pm{\rm i} k_{0j} a_z)}$, respectively. Therefore, one can
present the coefficients in the off-diagonal components of $V_m$
as \cite{Fu93,Fu94}
\begin{equation} \label{exp}
\Lambda_m = \lambda_m {\rm e}^{- 2 {\rm i} k_0 z_m}\:,\: P_m = p_m
{\rm e}^{- 2 {\rm i} k_0 z_m}\:,\: P'_m = p'_m {\rm e}^{- 2 {\rm i}
k_0 z_m}\:,
\end{equation}
where the complex coefficients $\lambda_m, p_m, p'_m$ are
independent of the interface position. In the following we assume
the origin $z=0$ to lie in the QW center.

The structure B/A/B is invariant under the mirror rotation
operation ${\cal S}_4$ with the transformation center at $z=0$, if
the number $N$ of atomic planes in the layer A is odd, and under
the space inversion operation $i$, if $N$ is
even.\cite{Golub_Ivchenko} This symmetry property allows one to
establish the relations between the coefficients in
Eq.~(\ref{2valley}) for the left- and right-hand side interfaces.
Since each of the operations results in the reciprocal
transformation $\psi_{ {\bm k}_{0j} }({\bm r}) \leftrightarrow
\psi_{- {\bm k}_{0j} }({\bm r})$ one has
\[
\left[ \begin{array}{cc} 0 & \lambda_{\rm L} \\ \lambda^*_{\rm L}
& 0 \end{array} \right] = \left[ \begin{array}{cc} 0 & 1 \\
1 & 0 \end{array} \right] \left[
\begin{array}{cc} 0 & \lambda_{\rm R} \\ \lambda^*_{\rm R} & 0
\end{array} \right] \left[ \begin{array}{cc} 0 & 1 \\ 1
& 0 \end{array} \right]
\]
or, equivalently, $\lambda_{\rm L} = \lambda^*_{\rm R}$. Taking
into account that, under the mirror-rotation operation ${\cal
S}_4$, the C$_{2v}$-group invariants $h({\bm k})$ and $h'({\bm
k})$ transform, respectively, into $h({\bm k})$ and $- h'({\bm
k})$ while, under the space inversion $i$, both $h({\bm k})$ and
$h'({\bm k})$ change their sign, we also obtain the relations
\begin{equation} \label{srsl}
S_{\rm R} = S_{\rm L} \:,\: S'_{\rm R} = - S'_{\rm L} \:,\: p_{\rm
R} = p^*_{\rm L}\:,\: p'_{\rm R} = - p'^*_{\rm L} \hspace{2.7 cm}
(\mbox{odd $N$})
\end{equation}
and
\begin{equation} \label{srslm}
S_{\rm R} = - S_{\rm L}\:,\: S'_{\rm R} = - S'_{\rm L}\:,\: p_{\rm
R} = - p^*_{\rm L} \:,\: p'_{\rm R} = - p'^*_{\rm L} \hspace{2.4
cm} (\mbox{even $N$}) \:.
\end{equation}
Hereafter we use the notations $\lambda, S, S', p, p'$ instead of
$\lambda_{\rm R}, S_{\rm R}, S'_{\rm R}, p_{\rm R}, p'_{\rm R}$.
By using Eqs.~(\ref{srsl}) and (\ref{srslm}) we can reduce the
components in the matrix (\ref{2valley}) to
\begin{equation} \label{oddn}
V_{{\rm R}, 11} = V_{{\rm R}, 22} = S h({\bm k}) + S' h'({\bm
k})\:,\: V_{{\rm L}, 11} = V_{{\rm L}, 22} = S h({\bm k}) - S'
h'({\bm k})\:,
\end{equation}
\[
V_{{\rm R}, 12} = V^*_{{\rm R}, 21} = {\rm e}^{- {\rm i} k_0 L}
[\lambda I + p\ h({\bm k}) + p' h'({\bm k})]\:,
\]
\[
V_{{\rm L}, 12} = V^*_{{\rm L}, 21} = {\rm e}^{{\rm i} k_0 L}
[\lambda^* I + p^* h({\bm k}) - p'^* h'({\bm k})]
\]
if $N$ is odd and to
\begin{equation} \label{evenn}
V_{{\rm R}, 11} = V_{{\rm R}, 22} = S h({\bm k}) + S' h'({\bm
k})\:,\: V_{{\rm L}, 11} = V_{{\rm L}, 22} = - S h({\bm k}) - S'
h'({\bm k})\:,
\end{equation}
\[
V_{{\rm R}, 12} = V^*_{{\rm R}, 21} = {\rm e}^{- {\rm i} k_0 L}
[\lambda I + p\ h({\bm k}) + p' h'({\bm k})]\:,
\]
\[
V_{{\rm L}, 12} = V^*_{{\rm L}, 21} = {\rm e}^{{\rm i} k_0 L}
[\lambda^* I - p^* h({\bm k}) - p'^* h'({\bm k})]
\]
if $N$ is even. Here $L=z_{\rm R} - z_{\rm L}$ is the QW width, it
is given by $L = N a_0/4$ with $a_0$ being the zinc-blende lattice
constant.

Equations (\ref{oddn}) and (\ref{evenn}) present the results of
the extended envelope-function method and yield relations between
coefficients in the matrices $V_L$ and $V_R$ for macroscopically
symmetric QWs.

If the barriers are grown from different materials B and C then
the coefficients in Eqs.~(\ref{oddn}), (\ref{evenn}) should be
labeled by the interface index, C/A or B/A, e.g., $S({\rm C/A})$
and $p'({\rm B/A})$. The different positions of the extremum
points ${\bm k}_{0j}$ are easily taken into account by replacing
$\varphi_1(z; j)$ and $\varphi_2(z; j)$ ($j =$ B, C) in
Eq.~(\ref{psi}) and in the boundary conditions (\ref{BC}) by
$$\tilde{\varphi}_1(z; j) = {\rm e}^{{\rm i} (k_{0j} - k_{0{\rm A}}) z_j}
\varphi_1(z; j)\:,\: \tilde{\varphi}_2(z; j) = {\rm e}^{- {\rm i}
(k_{0j} - k_{0{\rm A}}) z_j}  \varphi_2(z; j)\:,
$$
where $z_j$ is the coordinate of the interface between the layers
A and $j=$ C or B. This replacement allows to retain the form of
the perturbation ${\cal H}'$ defined by Eqs.~(\ref{h1}),
(\ref{2valley}) and (\ref{exp}).
\subsection{Valley-orbit splitting}
\begin{figure*}
  \centering
    \includegraphics[width=.35\textwidth]{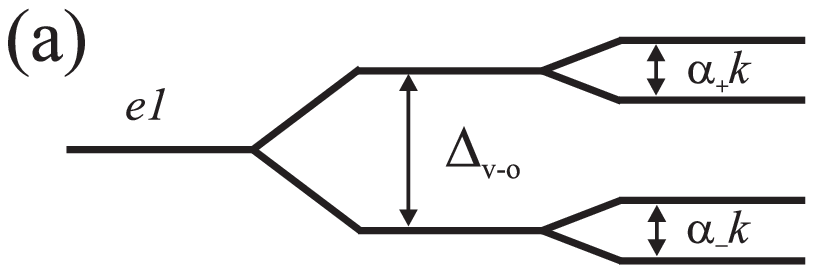}
    \hspace{0.2cm}
    \includegraphics[width=.35\textwidth]{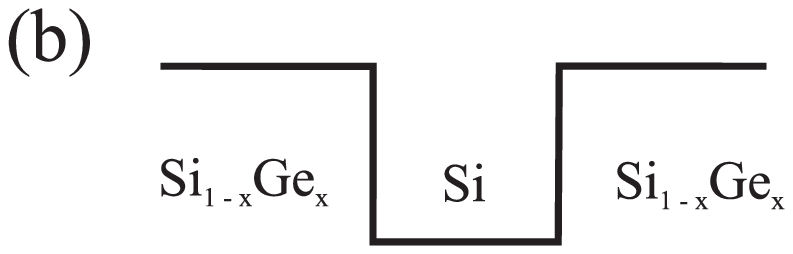}
  \caption{
Schematic representation of (a) hierarchy of the $e1$ subband splittings
and (b) the Si/Si$_{1-x}$Ge$_x$ structure under consideration.
We remind that, in this structure, the conduction band offset is mostly
determined by the strain; 
}\label{f_scheme}
\end{figure*}

The numerical calculations presented in the following sections
confirm the hierarchy
\begin{equation}\label{hierarchy}
E_X - E({\bm k}_0) \gg E_{e1} \gg \Delta_{\rm v\mbox{-}o} \gg
\Delta_{\rm spin}\equiv \alpha_{\pm} k\:
\end{equation}
illustrated by Fig.~\ref{f_scheme}a. Here $E({\bm k}_0)$ and $E_X$
are the conduction-band energies at the extremum point ${\bm k}_0$
and the $X$ point in the bulk material A, $E_{e1}$ is the
quantum-confinement energy for the lowest conduction subband,
$\Delta_{\rm v\mbox{-}o}$ and $\Delta_{\rm spin}$ are the
valley-orbit and spin splitting of the $e1$-subband states.
Therefore, we can line up the discussion in series starting from
the quantum confinement, turning then to the valley-orbit
splitting and finally to the spin splitting. As above we start
from the analysis of the symmetric structure B/A/B shown
schematically in Fig.~\ref{f_scheme}b and then generalize the
results on asymmetric structures with different barriers B and C.

For eigenstates of the zero-approximation Hamiltonian ${\cal H}_0$
the inter-valley mixing is absent and the envelope functions
referred to the first and second (001)-valleys form identical
sets. In particular, for the $e1$ subband states in the B/A/B
structure, the envelope has the standard form
\begin{equation} \label{211-7}
\chi(z)= c \left \{\begin{array}{c} \mbox{} \hspace{19
mm}\cos{qz}\:,\hspace{23
mm}\mbox{if} \hspace{5 mm} \vert z \vert \leq L/2 \:,\\
\cos{(qL/2)} \exp{[- \mbox{\ae} (\vert z \vert -
L/2)]}\:,\hspace{5 mm}\mbox{if} \hspace{5 mm} \vert z \vert \geq
L/2 \:.
\end{array} \right.
\end{equation}
Here $q = [ 2 m_{l}({\rm A}) E_{e1}/\hbar^2]^{1/2}$, $\mbox{\ae} =
[ 2 m_{l}({\rm B}) (V - E_{e1})/\hbar^2]^{1/2}$ and $c$ is the
normalization factor. The size-quantization energy $E_{e1}$
satisfies the transcendental equation $\tan{(q L/2)} = (\mbox{\ae}
/ q) [m_l({\rm A}) / m_l({\rm B})]$.

Now we switch on the inter-valley mixing taking into account
zero-${\bm k}$ terms in Eqs.~(\ref{oddn}) and (\ref{evenn})
proportional to $\lambda$ and $\lambda^*$. According to
Eqs.~(\ref{h1}),~(\ref{oddn}) and~(\ref{evenn}) the matrix element
of the inter-valley coupling is given by
\begin{equation} \label{matel}
M_{1, s ; 2, s'} =  \left| \chi \left(L/2 \right) \right| ^2
(V_{{\rm R},12}+V_{{\rm L},12}) =2
\left| \chi \left(L/2 \right) \right| ^2 |\lambda|
\cos{(k_0L - \phi_{\lambda})} \delta_{ss'}\:,
\end{equation}
where $|\lambda|$ and $\phi_{\lambda}$ are the modulus and the
phase of $\lambda$, and $s, s' = \pm 1/2$ are the electron spin
indices. Thus, the energies of the split $e1$ states at
$k_x=k_y=0$ are
\begin{equation} \label{e1ener}
E_{e1, \pm } = \pm 2\left| \chi \left(L/2 \right) \right| ^2
\cdot |\lambda  \cos{(k_0L - \phi_{\lambda})}|\:,
\end{equation}
and the envelopes are
\begin{equation} \label{e1envel}
\varphi_1(z; e1, \pm) = \pm \eta \varphi_2(z; e1, \pm) =
\chi(z) / \sqrt{2} \:,
\end{equation}
where $\chi(z)$ is defined in Eq.~(\ref{211-7}) and $\eta = {\rm
sign} \{ \cos{(k_0L - \phi_{\lambda})} \}$. Therefore, the parity
of the lower state $|e1, - \rangle$ (with respect to the operation
${\cal S}_4$ if $N$ is odd and $i$ if $N$ is even) follows the
sign of $\eta$ and reverses with the reversal of $\eta$. Equation
\eqref{e1ener} expresses an oscillating character of
$\Delta_{\rm{v-o}}(L)$ in terms of the envelope function method.
It will be shown in Sec.~III that this agrees with the
tight-binding numerical results.

For an asymmetric C/A/B structure, the inter-valley matrix element and
$e1$-subband energies are generalized to
\begin{equation} \label{asymm}
M_{1, s ; 2, s'} = \left| \chi \left(L/2 \right) \right| ^2 \left(
\lambda_{\rm R} {\rm e}^{ - {\rm i} k_{0A} L} + \lambda_{\rm L}
{\rm e}^{ {\rm i} k_{0A} L} \right) \delta_{ss'}\:,
\end{equation}
\[
E_{e1, \pm } = \pm \left| \chi \left(L/2 \right) \right| ^2\
 [\ |\lambda_{\rm R}|^2 +
|\lambda_{\rm L}|^2 + 2 |\lambda_{\rm R} \lambda_{\rm L}| \cos{(2
k_{0A} L + \phi_{\lambda_{\rm L}} - \phi_{\lambda_{\rm R}})}\
]^{1/2}\:,
\]
where $\phi_{\lambda_m}$ is the phase of $\lambda_m$ ($m = {\rm
R}, {\rm L}$).
\subsection{Spin-orbit splitting}
The next step is to take into account spin-dependent terms in
$V_m$. Since the symmetry forbids spin-splitting of the electron
states in the B/A/B system with an even number of atomic planes in
the A layer, we set $N$ to be odd. Then the inter-valley mixing is
described by the matrix elements
\begin{equation} \label{matspin}
M_{1, s ; 2, s'}
= 2 \left| \chi \left(L/2 \right) \right| ^2 [\ |\lambda|
\cos{(k_0L - \phi_{\lambda})}
\delta_{ss'} + |p|\cos{(k_0L - \phi_p)}\ h_{ss'}({\bm k}) - {\rm
i} |p'|\sin{(k_0L - \phi_{p'})}\ h'_{ss'}({\bm k}) \ ]\:,
\end{equation}
where $\phi_p, \phi_{p'}$ are the phases of $p$ and $p'$. Assuming
the valley-orbit splitting to exceed the spin-orbit splitting we
are able to rewrite the Hamiltonian in the basis (\ref{e1envel})
and obtain the following 2$\times$2 spin-dependent effective
Hamiltonians in the subbands $(e1, \pm)$
\begin{equation}
{\cal H}'({\bm k}; e1, \pm) = \left| \chi \left( L/2\right)
\right| ^2 \ \left[ V_{{\rm L}, 11}+V_{{\rm R}, 11}\pm \eta\ {\rm
Re}(V_{{\rm L}, 12}+V_{{\rm R}, 12}) \right] \:,
\end{equation}
and finally
\begin{equation} \label{alpha}
{\cal H}({\bm k}; e1, \pm) = E_{e1,\pm} + \alpha_{\pm} h({\bm
k})\:,
\end{equation}
where the coefficients in the linear-${\bm k}$ term are given by
\begin{equation} \label{a+-}
\alpha_{\pm} = 2\left| \chi \left(L/2 \right) \right| ^2
[\ S \pm |p| \eta \cos{(k_0L -
\phi_p)}\ ]\:.
\end{equation}
While deriving Eq.~(\ref{a+-}) we took into account both the
intra- and inter-valley contributions to $V_m$ and retained only
the terms up to the first order in $S$ and $p$. In agreement with
the symmetry arguments, neither the $S'$-dependent nor
$p'$-dependent contributions to $V_m$ give rise to linear-${\bm
k}$ terms. Note that, for the sake of completeness, in addition to
the linear-${\bm k}$ terms one can include in the right-hand side
of Eq.~(\ref{a+-}) a spin-independent quadratic-${\bm k}$ term
$\hbar^2 {\bm k}^2/2 m_{\parallel}$. Here $m_{\parallel}^{-1} =
\langle e1 | m_{\parallel}^{-1}(z) | e1 \rangle$ and the angle
brackets mean averaging over the $e1$ state defined in
Eq.~(\ref{211-7}).

In addition to a smoothly decreasing term in $\alpha_{\pm}$
predicted in Ref.~[\onlinecite{Golub_Ivchenko}], Eq.~\eqref{a+-}
contains an oscillating term. The reason for the oscillations is
mixing of valley states at the QW interfaces. Tight-binding
calculations presented below show that $\left|p\right| > S$, i.e.,
the oscillating part of $\alpha_{\pm}$ is dominating.

For an asymmetric structure C/A/B, the linear-${\bm k}$
contribution to the Hamiltonian ${\cal H}({\bm k}; e1, \pm)$ takes
the form
\[
{\cal H}^{(1)}({\bm k}; e1, \pm) = \alpha_{\pm} h({\bm k}) +
\beta_{\pm} h'({\bm k})
\]
with
\begin{equation} \label{a+-a}
\alpha_{\pm} = \left| \chi \left(L/2 \right) \right| ^2 [\
S_{\rm R} + S_{\rm L}  \pm  {\rm Re} \{ {\rm e}^{- {\rm i} \xi}
(p_{\rm R} {\rm e}^{- {\rm i} k_0 L} + p_{\rm L} {\rm e}^{{\rm i}
k_0 L}) \} \ ]\:,
\end{equation}
\[
\beta_{\pm} = \left| \chi \left(L/2 \right) \right|^2 [\
S'_{\rm R} + S'_{\rm L} \pm {\rm Re} \{ {\rm e}^{- {\rm i} \xi}
(p'_{\rm R} {\rm e}^{- {\rm i} k_0 L} + p'_{\rm L} {\rm e}^{{\rm
i} k_0 L}) \} \ ]\:,
\]
and $\xi = \arg{ \{ \lambda_{\rm R} {\rm e}^{- {\rm i} k_0 L} + \lambda_{\rm L} {\rm e}^{{\rm i} k_0 L} \}}$. For the symmetric
structure, ${\rm e}^{{\rm i} \xi} = \eta = \pm 1$.

\section{Tight-binding calculations and discussion}\label{s_results_n_discussion}
In order to estimate values of spin and valley-orbit splittings
we have performed calculations of the electron dispersion in the
$e1$ conduction subband by using one of the empirical tight-binding
models. More precisely, we have fixed on the nearest-neighbor
$sp^3s^*$ tight-binding model optimized for the conduction
band.\cite{Boikin00} This model is a reasonable compromise between
the numerical load and the accuracy of representation of the band
structure. It is capable to reproduce the indirect gap although
shifts the position of the conduction band minimum from the
experimentally measured point $k_0 = 0.85 \times 2 \pi/a_0$ to
the point $k_0 = 0.62 \times 2 \pi/a_0$. Note that a value of
$k_0$ is hardly reproduced even in the more sophisticated methods,
namely, the second-nearest neighbor $sp^3s^*$ and nearest neighbor
$sp^3d^5s^*$ tight-binding models,\cite{Boykin,Boykin_d} leading
to $k_0 a_0/ 2 \pi = 0.758$ and 0.813, respectively. The
applicability of the $sp^3s^*$ model is confirmed by the fact, see
below, that values of the valley-orbit splitting $\Delta_{\rm{v-o}}$
calculated in this work and by using the $sp^3d^5s^*$
model~\cite{Boykin_splitting} are of the same order of magnitude.

The empirical $sp^3s^*$ tight-binding method was previously
applied for calculation of the spin splitting in bulk GaAs and
GaAs-based QWs.\cite{Santos} The linear-in-${\bf k}$ splitting in
a QW was compared with the cubic spin splitting in bulk GaAs where
the component $k_z$ was replaced by $\pi/d_{\rm GaAs}$ with
$d_{\rm GaAs}$ being the width of the GaAs layer. The agreement
was obtained after replacing $d_{\rm GaAs}$ by an effective value
$d_{\rm GaAs}^{\rm eff}$ and adjusting the coefficient $\gamma$ in
the cubic-in-${\bf k}$ contribution to the electron effective
Hamiltonian ${\cal H}^{(1)}({\bm k})$. The need in the
introduction of the effective parameters $d_{\rm GaAs}^{\rm eff}$
and $\gamma^{\rm eff}$ can be related to an additional
contribution to ${\cal H}^{(1)}({\bm k})$ coming from the reduced
symmetry of interfaces, or in other words from the anisotropic
orientation of interface bonds~\cite{Voisin}. In contrast to the
zinc-blende-lattice heterostructures, in diamond-lattice QWs the
Hamiltonian ${\cal H}^{(1)}({\bm k})$ has no bulk inversion
asymmetry term proportional to $\gamma$ and is contributed only by
the interface inversion asymmetry term described by the
coefficient $\alpha$ in Eq.~(\ref{1}).~\cite{Golub_Ivchenko}

In the tight-binding method the electron Hamiltonian is presented
by a set of matrix elements taken between atomic orbitals. If a
heterostructure is grown from diamond-like semiconductors along
the [001] principal axis one can write the tight-binding
free-electron wave function
\begin{equation}
\psi = \sum_{n, \nu} { c_{n, \nu} \Phi_{n,\nu,\bm{k}}(\bm{r}) }
\end{equation}
in terms of planar orbitals
\begin{equation} \label{planar}
\Phi_{n, \nu, \bm{k}}(\bm{r}) = \sum_m {\rm e}^{ {\bm i} {\bm k} \cdot {\bm
r}_m } \Phi_{\nu} (\bm{r}-\bm{r}_m)\:.
\end{equation}
Here $n = 0, \pm 1,\pm 2...$ is the number of atomic planes
perpendicular to the growth direction $z
\parallel [001]$, $\Phi_{\nu}$ is the orthogonalized atomic
orbital with $\nu$ being the orbital index, the index $m$
enumerates atoms in the $n$-th atomic plane, ${\bm r}_m$ is the
position of the $m$th atom in this plane, in particular, $z_m = n
a_0 / 4$, ${\bm k}$ is the two-dimensional in-plane electron wave
vector. The index $\nu$ runs through $2N$ values where $N$ is the
number of orbitals taken into consideration and the factor is due
to electron spin. For convenience we use below the Cartesian
coordinate system $x'
\parallel [1 \bar{1} 0], y' \parallel [110], z \parallel [001]$.
In the nearest-neighbor approximation we obtain the following set
of equations
\begin{eqnarray}
&&\mbox{} \hspace{3 mm} \hat{U}_{y'}^{\dag}(2l) C_{2l - 1} \hspace{2 mm}
+ \hspace{2 mm} \hat{E}_0(2l) C_{2l} \hspace{1 mm} + \hspace{2
mm}\hat{U}_{x'}(2l) C_{2l + 1} \hspace{2 mm} = E C_{2l}\:, \\
&&\hat{U}_{x'}^{\dag}(2l - 1) C_{2l - 2} + \hat{E}_0(2l - 1) C_{2l - 1} +
\hat{U}_{y'}(2l - 1) C_{2l} = E C_{2l - 1} \nonumber
\end{eqnarray}
for the vectors $C_n$ containing $2N$ components $c_{n, \nu}$.
Here $\hat{E}_0(n)$ is ${\bm k}$-independent diagonal matrices,
$\hat{U}_{x'}$ and $\hat{U}_{y'}$ are $k_{x'}$ and $k_{y'}$
dependent matrices. The diamond lattice has two atoms per unit
cell and can be represented as two face-centered cubic sublattices
shifted with respect to each other by $\sqrt{3} a_0/4$ along the
[111] direction. The atomic planes with even $n = 2l$ and odd $n =
2l + 1$ ($l=0,\pm1...$) belong to the different sublattices and
differ in the direction of chemical bonds. As compared with the
pair of planes $2l$ and $2l + 1$, the orientation of chemical
bonds between atoms in the planes $2l - 1$ and $2l$ are rotated
around the axis $z$ by 90$^{\circ}$. For brevity we omit here the
detailed form of matrices $\hat{U}_{x',y'}(n)$; for
$k_{x'}=k_{y'}=0$ these matrices can be readily obtained from
those for the zinc-blende-based heterostructures given in
Ref.~[\onlinecite{Our_PRB}]. The matrices $\hat{E}_0$,
$\hat{U}_{x',y'}$ are formed by the tight-binding parameters,
which are usually extracted from fitting bulk-material band
structure to experimental one. The tight-binding parameters for Si
and Ge are listed in Table~\ref{tbl1}. The diagonal energies are
referred to the valence band top of each material.
\begin{table}
\caption{Tight-binding parameters used in the
calculations in eV.}\label{tbl1}
\begin{ruledtabular}
\begin{tabular}{c|ccc|ccccc|c} 
     & $E_{s}$  & $E_{p}$ & $E_{s^*}$ & $V_{ss}$ & $V_{xx}$ & $V_{xy}$ & $V_{sp}$ & $V_{s^*p}$ & $\Delta$ \\ \hline
Si   & -3.65866 & 1.67889 & 3.87576 & -7.97142   & 1.69558  & 23.32410 & 8.87467  & 5.41174 & 0.045 \\
Ge   & -5.88    &  1.61   & 6.39    &  -6.78     &  1.61    & 4.90     & 5.4649   & 5.2191  &  0.30 \\
\end{tabular}
\end{ruledtabular}
\end{table}
The parameters for Si were taken from
Ref.~[\onlinecite{Boikin00}]; those for Ge are not so critical for
the purpose of this work, we collected them from
Ref.~[\onlinecite{Vogl}] and added a value of 0.30 eV for the
spin-orbit splitting of the $p$ orbitals~[\onlinecite{Boykin_d}].
For SiGe alloys, we have used the virtual crystal approximation
and the linear interpolation of the tight-binding parameters. The
strain was taken into account only by shifting the diagonal
energies $E_{0, \nu}$ in Si or Ge by the same value, the
strain-induced splitting of the $p$-orbital states was ignored.
The shift of diagonal energies for the barrier material is equal
to $\Delta E_c - \Delta E_g$, where $\Delta E_g$ is the difference
in the band gaps of the well and barrier bulk materials and
$\Delta E_c$ is the conduction-band offset. For a
Si$_{1-x}$Ge$_x$/Si/Si$_{1-x}$Ge$_x$ QW structure with the
strained Si layer and the Ge content $x = 0.25$ we used a value of
$\Delta E_c = 0.15$ eV relying on
Refs.~[\onlinecite{Rieger,Schaffler,Wilamowski,Boykin_splitting}].

\begin{figure}
  \centering
    \includegraphics[width=.7\textwidth]{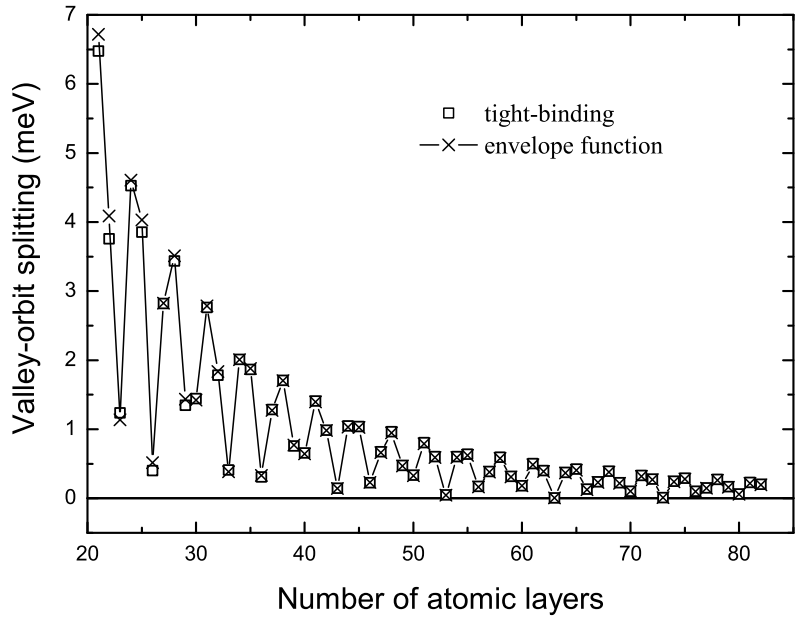}
  \caption{Valley-orbit splitting $\Delta_{\rm{v-o}}$ in
  Si$_{1-x}$Ge$_x$/Si/Si$_{1-x}$Ge$_x$ ($x$ = 0.25)
QW versus the number of Si mono atomic layers. Analytical results
shown by crosses calculated using Eq.~\eqref{e1ener} with
$|\lambda|=385$~meV$\cdot$~\AA,
$\phi_\lambda=0.3\pi$.}\label{f_vo}
\end{figure}

Squares in Fig.~\ref{f_vo} show results of tight-binding
calculations of the valley-orbit splitting $\Delta_{\rm
v\mbox{-}o}$ in symmetrical
Si$_{0.75}$Ge$_{0.25}$/Si/Si$_{0.75}$Ge$_{0.25}$ QWs as a function
of the number $N$ of Si atomic planes sandwiched between the thick
barriers Si$_{0.75}$Ge$_{0.25}$. The valley-orbit splitting
exhibits pronounced oscillations with the increasing QW width, in
agreement with
Ref.~[\onlinecite{Ohkawa,Boykin_splitting,Boykin_split2}]. The
oscillation periods in Fig.~\ref{f_vo} of the present work and in
Fig.~3 of Ref.~[\onlinecite{Boykin_split2}] vary considerably due
to the difference in values of ${\bm k}_0$ obtained in the
$sp^3s^*$ model used here and the $sp^3d^5s^*$ model. However, the
splittings $\Delta_{\rm v\mbox{-}o}$ are of the same order of
magnitude, e.g., at $N$ $\approx$ 60 the oscillation amplitudes
differ only by a factor of $\sim$2 which can be explained by the
obvious sensitivity of $\Delta_{\rm v\mbox{-}o}$ to the model
used.

Crosses in Fig.~\ref{f_vo} represent the calculation of
$\Delta_{\rm v\mbox{-}o}$ in the envelope-function approximation,
Eq.~(\ref{e1ener}), with $k_0 = 0.62 \times 2 \pi/a_0$. While
calculating the electron envelope function at the interface,
$\chi(L/2)$, we used values of $V=150$~meV for the conduction-band
offset and of 0.907$m_0$ ($m_0$ is the free electron mass) for the
longitudinal effective mass $m_l({\rm A})$ as obtained in the
$sp^3s^*$ tight-binding model optimized for the conduction
band,\cite{Boikin00} and, for simplicity, took $m_l({\rm B})$
equal to $m_l({\rm A})$. The modulus $|\lambda|$ and the phase
$\phi_{\lambda}$ were considered in Eq.~(\ref{e1ener}) as
adjustable parameters. Their best fit values turned out to be
$|\lambda|=385$~meV$\cdot$~\AA, $\phi_\lambda=0.3\pi$. It is seen
from Fig.~\ref{f_vo} that the simple analytical theory developed
in Sec.~II is in complete agreement with the results of more
sophisticated tight-binding calculations.

\begin{figure}
  \centering
    \includegraphics[width=.7\textwidth]{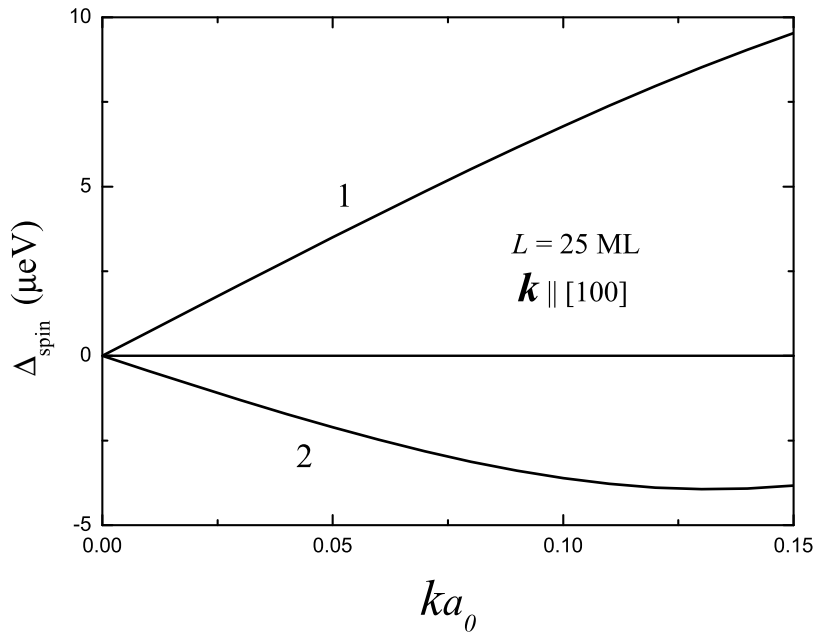}
\caption{Spin-splitting of the valley-orbit split subbands in a
Si$_{0.75}$Ge$_{0.25}$/(Si)$_N$/Si$_{0.75}$Ge$_{0.25}$ QW with $N=
25$ as a function of the in-plane wave vector for ${\bm k}
\parallel [100]$. Curves 1 and 2 correspond to the subbands
$E_{e1,+}$ and $E_{e1,-}$, respectively. The definition of the
sign of $\Delta_{\rm spin}$ is given in the text.}\label{f_so_k}
\end{figure}

\begin{figure}
  \centering
    \includegraphics[width=.7\textwidth]{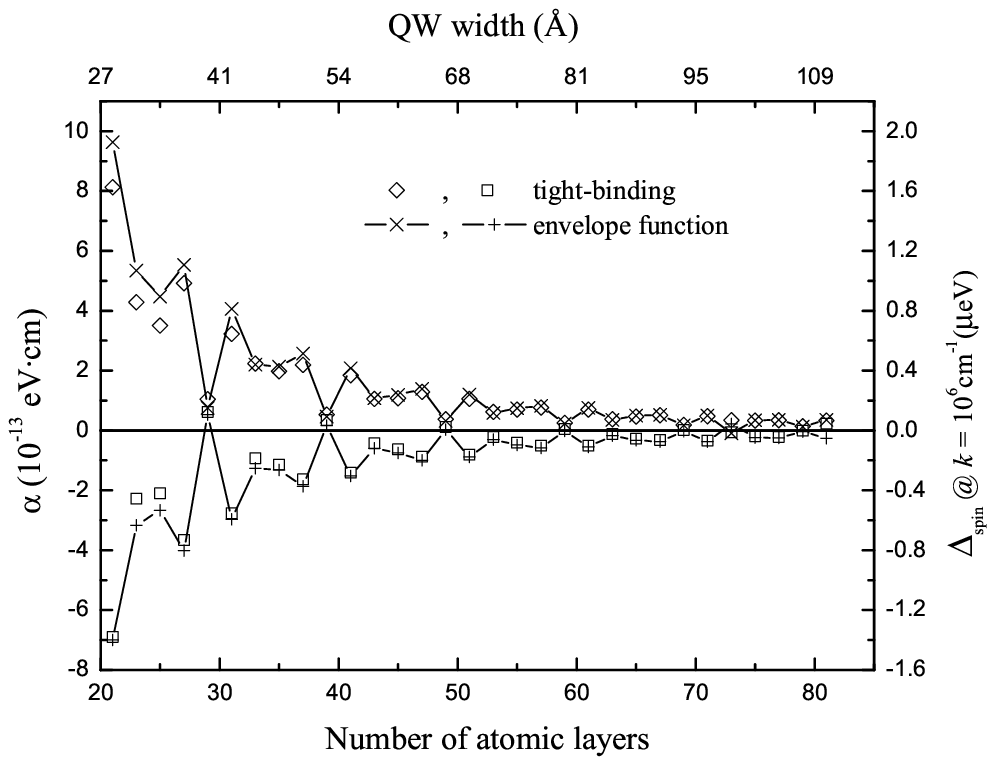}
\caption{Spin-splitting constant $\alpha$ in Eqs.~(\ref{alpha}),
(\ref{dakdis}) versus the QW width determined by the number of Si
monoatomic layers (odd $N$ are taken in consideration only). The
spin splitting of the lower subband $E_{e1-}$ is shown by diamonds
(tight-binding calculation) and x-shaped crosses (envelope
function approximation), those for the upper subband $E_{e1+}$ are
shown by squares and conventional crosses.}\label{f_so}
\end{figure}

In Figs.~\ref{f_so_k}, \ref{f_so} the
spin-orbit splitting for the two valley-orbit subbands $E_{e1,-}$
and $E_{e1,+}$ are presented. This is the first calculation of the
spin-splitting, no previous theoretical estimations are available
in order to compare with. We define the splitting $\Delta_{\rm
spin}$ in Fig.~\ref{f_so} as the energy difference between the
states with the spin parallel and antiparallel to the $x$ axis.
Then if the antiparallel state lies higher the sign of
$\Delta_{\rm spin}$ is negative as in case of the upper
valley-orbit split subband $E_{e1,+}$, see curve 2 in
Fig.~\ref{f_so_k}. The calculation shows that, up to $k \leq
10^6$~cm$^{-1}$, the linear dependence
\begin{equation} \label{dakdis}
\Delta_{\rm spin}(k) = \alpha_\pm k
\end{equation}
holds, in agreement with the Hamiltonian (\ref{a+-}). It is the
variation of $\alpha_{\pm}$ with odd $N$ which is shown in
Fig.~\ref{f_so}. As one can see from to
Figs.~\ref{f_vo}--\ref{f_so} the valley-orbit and spin splittings
are conveniently presented in the meV and $\mu$eV scales
confirming our assumption (\ref{hierarchy}).

Figure~\ref{f_so} shows that the spin splitting $\Delta_{\rm
spin}$ is an oscillating function of the QW width. This
demonstrates that the inter-valley spin-dependent mixing at the
interfaces prevails over the intra-valley contribution to
$\alpha_{\pm}$. Squares and diamonds in Fig.~\ref{f_so} show
results of tight-binding calculation. The spin splitting is
plotted only for odd number of Si monoatomic planes because, for
even $N$, $\Delta_{\rm spin}$ in the symmetric structures
vanishes. Conventional and x-shaped crosses are obtained as the
best fit using Eq.~(\ref{a+-}) and choosing the same values for
$k_0$ and $\phi_{\lambda}$ as in Fig.~\ref{f_vo} and the
additional adjustable parameters $|p| = 0.53 \cdot
10^{-5}$~eV$\cdot$cm$^2$, $\phi_p = 0.55$, $S = 0.15 |p|$.

Now we compare the value of $\alpha_-$ estimated in this work with
that extracted by Wilamowski et al.~\cite{Wilamowski} from
spin-resonance measurements in a Si/Si$_{1-x}$Ge$_x$ QW structure
with $x=0.25$. Note that the value $\alpha_- = 0.55\cdot 10^{-12}$
eV$\cdot$\AA\mbox{} presented in this reference for a 120\AA-thick
QW should be decreased by a factor of $1.6$, i.e., in fact
$\alpha_- = 0.34\cdot 10^{-12}$ eV$\cdot$\AA\mbox{}, see
Ref.~[\onlinecite{Glazov}]. Our estimation of $\alpha_-$ gives a
value smaller by a factor $\sim6$. This means that in the sample
studied in Ref.~[\onlinecite{Wilamowski}] the Rashba (or
structure-inversion asymmetry\cite{Ivchenko}) contribution to the
spin splitting dominates over the intrinsic contribution
considered here. Nevertheless, the experimental value of the spin
splitting is not so far from the limit for a perfect QW structure.
\section{Conclusion}
The $sp^3s^*$ tight-binding model has been developed in order to
calculate the electron dispersion in heterostructures grown from
multivalley semiconductors with the diamond lattice, particularly,
in the Si/SiGe structures. The model allows one to estimate the
orbit-valley and spin-orbit splittings of the electron
quantum-confined states in the ground subband. In the employed
tight-binding model, the spin-orbit splitting is mostly determined
by the spin-dependent orbit-valley mixing at the interfaces. For
this reason the coefficients $\alpha_{\pm}$ describing the
linear-in-${\bm k}$ splitting are strongly oscillating functions
of the odd number, $N$, of the Si monoatomic layers.

In addition to the numerical calculations, an envelope-function
approximation has been extended to take account of spin-dependent
reflection of an electronic wave at the interface and
interface-induced intervalley mixing. The dependencies of the
valley-orbit and spin-orbit splittings upon the number of Si
atomic planes calculated in the tight-binding microscopic model
are successfully reproduced by using simple analytical equations
derived in the envelope-function theory and fitting the parameters
that enter into these equations. It follows then that the
envelope-function approach can be applied as well for the
description of electron-subband splittings in a realistic Si/SiGe
structure.

\acknowledgments{This work was financially supported by
the RFBR, programmes of RAS, INTAS, ``Dynasty'' Foundation ---
ICFPM, and Russian President grant for young scientists}

\bibliography{bibliography}
\end{document}